\documentclass[a4paper,11pt]{scrartcl}

\usepackage[top=3cm,left=3cm,right=3cm,bottom=3cm]{geometry}
\usepackage[intlimits]{amsmath}
\usepackage{amsfonts,amssymb,amsthm}
\usepackage{mathtools}
\usepackage{hyperref}
\usepackage{enumerate}
\usepackage{braket}
\usepackage{graphicx}
\usepackage{graphbox}
\usepackage{booktabs}
\usepackage[section]{placeins}
\usepackage{caption}
\usepackage{subcaption}
\usepackage{color}
\usepackage{todonotes}

\usepackage{tikz}
\usepackage{mathtools}
\usetikzlibrary[graphs, graphs.standard]
\usetikzlibrary{positioning,chains,fit,shapes,calc,decorations.markings}
\usetikzlibrary{decorations.pathreplacing,calligraphy}

\tikzstyle{vertex}=[circle, draw, inner sep=0pt, minimum size=6pt]
\tikzstyle{marked}=[circle,draw,dotted,inner sep=0pt, minimum size=6pt]

\newcommand{\matrx}[1]{\left[\begin{matrix} #1 \end{matrix}\right]}

\hypersetup{pdfstartview=FitH}

\begin{document}

\title{Quantum state transfer on the complete bipartite graph}
\author{Raqueline A. M. Santos}
\date{\small{Center for Quantum Computer Science, Faculty of Computing, University of Latvia} \\ 
\small{Raina bulv. 19, Riga, LV-1586, Latvia}\\
\small{\texttt{rsantos@lu.lv}}}

\maketitle

\begin{abstract}
Previously it was shown that (almost) perfect state transfer can be achieved on the complete bipartite graph by a discrete-time coined quantum walk based algorithm when both the sender and receiver vertices are in the same partition of the graph and when the sender and receiver are in opposite partitions of the same size. By changing the coin operator, we analyze the state transfer problem and we show that it is still possible to achieve state transfer with high fidelity even when the sender and receiver are in different partitions with different sizes. Moreover, it is also possible to use an active switch approach using lackadaisical quantum walks where the marked vertex is switched between the sender and receiver during the algorithm.

\end{abstract}

%%
%% Introduction
%%
\section{Introduction}

Quantum walks have been a successful tool used in many algorithmic applications, specially in search algorithms~\cite{Portugal:2013}. Another related problem is the task of performing quantum state transfer~\cite{Bose:2003} between two vertices of a graph, often called sender and receiver. In this problem, we want to transfer with high probability a state localized initially on the sender vertex to the receiver vertex. 

Quantum state transfer has been investigated on different graphs by means of both discrete-time and continuous-time quantum walks. For example, quantum state transfer can be performed on the line~\cite{Yalcinkaya:2015, Shang:2019},  cycle~\cite{Kurzynski:2011,Kendon:2011,Barr:2014,Yalcinkaya:2015,Shang:2019}, lattices~\cite{Hein:2009, Zhan:2014}, star graphs and complete graphs with loops~\cite{Stefanak:2016}, regular graphs~\cite{Shang:2019, Zhan:2019} and different types of networks~\cite{Chen:2019,Cao:2019}.
In the case of continuous-time quantum walk, it was shown by~\cite{Chakraborty:2016} that perfect state transfer can be achieved for almost any graph.

 Closely related to the work in this paper, quantum state transfer by means of discrete-time coined quantum walk on the complete bipartite graph was analyzed in~\cite{Stefanak:2017}. They considered two marked vertices, the sender and receiver. They proved that (almost) perfect state transfer between these two vertices can be achieved when both the sender and receiver are in the same partition and when the sender and receiver belong to opposite partitions of the same size. Later, Štefaňák and Skoupý~\cite{Stefanak:2021} analyzed the same problem  on the complete $M$-partite graph with partitions of the same size, by using lackadaisical quantum walks. They showed
that when the sender and the receiver are in different partitions, the algorithm succeeds with fidelity approaching
unity for a large graph. And that is not the case when the sender and the receiver are in the same partition. They also proposed an algorithm with an active switch
based on a single marked vertex which achieves high fidelity for large graphs independent on the location of the sender and receiver vertices.

In this work, we analyze quantum state transfer between two vertices (sender and receiver) of the complete bipartite graph by means of discrete-time coined quantum walks. Previously, the same problem was analyzed by~\cite{Stefanak:2017}. By using a different coin operator from~\cite{Stefanak:2017}, we show that it is still possible to achieve state transfer with high fidelity even when the sender and receiver are in different partitions with different sizes. Next, we consider a similar approach to~\cite{Stefanak:2021} and we describe a quantum state transfer algorithm with active switch, which is based on one marked vertex. We show that for large graphs, independent on the location of the sender and receiver, the algorithm succeeds with fidelity approaching
unity. We simulate the algorithm and we find out that high fidelity can also be achieved for small graphs.  

The paper is organized as follows. In section~\ref{sec:QW}, we describe the coined quantum walk on the complete bipartite graph and how the state transfer algorithm works. We analyze state transfer when the sender and receiver belong to different partitions in section~\ref{sec:diff}, and when they are in the same partition in section~\ref{sec:same}. The quantum state transfer algorithm with active switch by means of lackadaisical quantum walks is described in section~\ref{sec:switch}. Summary and conclusions are presented in section~\ref{sec:conc}.

\section{Coined quantum walk on the complete bipartite graph}
\label{sec:QW}
\begin{figure}[!htb]
\centering
\begin{tikzpicture}[thick,
  every node/.style={draw,circle},
  fsnode/.style={fill=black!15},
  ssnode/.style={fill=black!15},
  every fit/.style={ellipse,draw,inner sep=-2pt,text width=2cm},
  -,shorten >= 0pt,shorten <= 0pt
]

%Braces
\begin{scope}[every node/.style={}]
\draw [decorate, 
    decoration = {calligraphic brace,
        raise=5pt,
        amplitude=5pt}] (-0.35,0.1) --  (0.35,0.1)
    node[pos=0.5,above=10pt,black]{$V_1$};
\draw [decorate, 
    decoration = {calligraphic brace,
        raise=5pt,
        amplitude=5pt}] (1.35,-0.4) --  (2.05,-0.4)
    node[pos=0.5,above=10pt,black]{$V_2$};
\end{scope}

% the vertices of U
\begin{scope}[start chain=going below,node distance=4mm]
\foreach \i in {1,2,...,4}
  \node[fsnode,on chain,minimum size=12pt,inner sep=0pt] (f\i) {};

\end{scope}

% the vertices of V
\begin{scope}[xshift=1.7cm,yshift=-0.5cm,start chain=going below,node distance=4mm]
\foreach \i in {1,2,...,3}
  \node[ssnode,on chain,minimum size=12pt,inner sep=0pt] (s\i) {};
\end{scope}

% the edges
\draw (f1) -- (s1);
\draw (f1) -- (s2);
\draw (f1) -- (s3);
\draw (f2) -- (s1);
\draw (f2) -- (s2);
\draw (f2) -- (s3);
\draw (f3) -- (s1);
\draw (f3) -- (s2);
\draw (f3) -- (s3);
\draw (f4) -- (s1);
\draw (f4) -- (s2);
\draw (f4) -- (s3);
\end{tikzpicture}

\caption{A complete bipartite graph with $N_1 = 4$ and $N_2=3$.}
\label{fig:bip}
\end{figure}
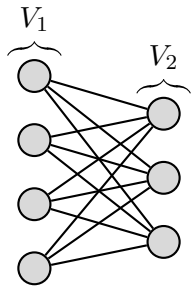

Let $\Gamma(V_1,V_2,E)$ be a complete bipartite graph with $N_1 = |V_1|$ and $N_2 = |V_2|$, as in figure~\ref{fig:bip}. 

Associated to the graph, we define the Hilbert space $\mathcal{H}^{2|E|}$ spanned by the basis states $\ket{v,u}$ which represents the particle located at vertex $v$ pointing to vertex $u$. 
A step of the quantum walk is obtained by applying the evolution operator
\begin{equation}\label{eq:U}
    U = SC.
\end{equation}
The coin operator $C$ is defined by
\begin{equation}
    C = \bigoplus_{v}C_v.
\end{equation}
Since the vertices can have different degree, $C$ is defined as a direct sum, so that $C_v$ acts locally at vertex $v$.
The operator $S$ is the flip-flop shift which acts as
\begin{equation}
    S\ket{v,u} = \ket{u,v},
    \label{eq:shift}
\end{equation}
that is, the particle will move from vertex $v$ to vertex $u$ after its application.

State transfer using discrete-time quantum walks can be realized by considering two marked vertices, the sender $s$ and the receiver $r$. In this paper, we use the Grover diffusion operator $G$ as the coin operator on the non-marked vertices, and we apply $-G$ to the marked ones. The Grover coin acts on a vertex as
\begin{equation}
    G_v = \ket{v}\bra{v}\otimes(2\ket{\mu_v}\bra{\mu_v}-I),
    \label{eq:grover}
\end{equation}
where
\begin{equation}
    \ket{\mu_v} = \frac{1}{\sqrt{d_v}}\sum_{u\sim v}\ket{u},
\end{equation}
and $d_v$ is the degree of vertex $v$.
Without loss of generality, the sender vertex is always located in the partition $V_1$. And we will consider both the cases when $r\in V_1$ and $r\in V_2$.
For simplicity, we will use the sub-index to identify which partition the vertex belongs to, that is, $v_j \in V_j, j\in\{1,2\}$. And we consider that the sum over all vertices of a partition $V_j$ is represented as $\sum_{v_j} \equiv \sum_{v_j\in V_j}$.

The idea of the state transfer algorithm is as follows. The initial state is localized at the sender vertex,
\begin{equation}
    \ket{\psi(0)} = \frac{1}{\sqrt{N_2}}\sum_{v_2}\ket{s,v_2}.
    \label{eq:psi0}
\end{equation}
We apply the evolution operator $t$ times. The state of the walker after $t$ steps is given by
\begin{equation}
    \ket{\psi(t)} = U^t\ket{\psi(0)}.
\end{equation}
Then, we analyze the walk towards the target state, $\ket{target}$, localized at the receiver vertex. The fidelity between the state at time $t$ and the target state is given by
\begin{equation}
    \mathcal{F}(t) = |\braket{\psi(t)|target}|^2.
\end{equation}
We expect that after a certain number of steps we can achieve state transfer with high fidelity. Following, we analyze the cases where the sender and receiver belong to different and to the same partitions.

\section{Sender and receiver in different partitions}
\label{sec:diff}
\begin{figure}[!htb]
\centering
\begin{tikzpicture}[thick,
  every node/.style={draw,circle},
  fsnode/.style={fill=black!15},
  ssnode/.style={fill=black!15},
  every fit/.style={ellipse,draw,inner sep=-2pt,text width=2cm},
  -,shorten >= 0pt,shorten <= 0pt
]

%Braces
\begin{scope}[every node/.style={}]
\draw [decorate, 
    decoration = {calligraphic brace,
        raise=5pt,
        amplitude=5pt}] (-0.35,0.1) --  (0.35,0.1)
    node[pos=0.5,above=10pt,black]{$V_1$};
\draw [decorate, 
    decoration = {calligraphic brace,
        raise=5pt,
        amplitude=5pt}] (1.35,-0.4) --  (2.05,-0.4)
    node[pos=0.5,above=10pt,black]{$V_2$};
\end{scope}

% the vertices of U
\begin{scope}[start chain=going below,node distance=4mm]
\node[double,fill=green!15,on chain,minimum size=12pt,inner sep=0pt] (f1) {\footnotesize{$s$}};
\foreach \i in {2,...,4}
  \node[fsnode,on chain,minimum size=12pt,inner sep=0pt] (f\i) {};

\end{scope}

% the vertices of V
\begin{scope}[xshift=1.7cm,yshift=-0.5cm,start chain=going below,node distance=4mm]
\node[double,fill=red!15,on chain,minimum size=12pt,inner sep=0pt] (s1) {\footnotesize{$r$}};
\foreach \i in {2,...,3}
  \node[ssnode,on chain,minimum size=12pt,inner sep=0pt] (s\i) {};
\end{scope}

% the edges
\draw (f1) -- (s1);
\draw (f1) -- (s2);
\draw (f1) -- (s3);
\draw (f2) -- (s1);
\draw (f2) -- (s2);
\draw (f2) -- (s3);
\draw (f3) -- (s1);
\draw (f3) -- (s2);
\draw (f3) -- (s3);
\draw (f4) -- (s1);
\draw (f4) -- (s2);
\draw (f4) -- (s3);
\end{tikzpicture}

\caption{A complete bipartite graph with the sender vertex $s$ and receiver vertex $r$ in different partitions. In this case, $N_1 = 4$ and $N_2=3$. }
\label{fig:bip-diff}
\end{figure}
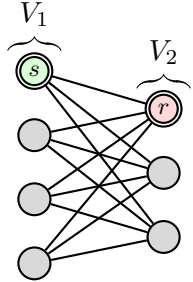

Consider that the sender and the receiver belong to different partitions, that is, the sender $s \in V_1$ and the receiver $r \in V_2$. See figure~\ref{fig:bip-diff} for an example. 

For this case, the coin operator is described as follows. We apply $-G$ to the marked vertices and $G$ to the non-marked, that is, $C_s = -G_s$, $C_{v_1} = G_{v_1} (v_1\neq s)$, $C_r = -G_r$, $C_{v_2} = G_{v_2}(v_2\neq r)$, where $G_v$ is given by~(\ref{eq:grover}). The evolution operator is $U_{s,r} = SC$.
%\begin{eqnarray}
%    C_{s} &=& -\frac{2}{N_2}\sum_{u_2, v_2}\ket{s,u_2}\bra{s,v_2} + \sum_{v_2}\ket{s,v_2}\bra{s,v_2},\\
%    C_{v_1} &=& \frac{2}{N_2}\sum_{u_2, v_2}\ket{v_1,u_2}\bra{v_1,v_2} - \sum_{v_2}\ket{v_1,v_2}\bra{v_1,v_2}, \quad\forall v_1\neq s.
%\end{eqnarray}
%And for the vertices in $V_2$,
%\begin{eqnarray}
%    C_{r} &=& -\frac{2}{N_1}\sum_{u_1, v_1}\ket{r,u_1}\bra{r,v_1} + \sum_{v_2}\ket{r,v_2}\bra{r,v_2},\\
%    C_{v_2} &=& \frac{2}{N_1}\sum_{u_1, v_1}\ket{v_2,u_1}\bra{v_2,v_1} - \sum_{v_1}\ket{v_2,v_1}\bra{v_2,v_1}, \quad\forall v_2\neq r.
%\end{eqnarray}

Due to the symmetry of the graph, it is possible to find an invariant subspace in which the quantum walk takes place. Since at the end of the algorithm we will be interested in the state of the walker being at  the receiver vertex $r$, we consider the states that contains vertices of $V_2$ pointing to $V_1$, that is,
\begin{eqnarray}
    \ket{\phi_1} &=& \ket{r,s},\\
    \ket{\phi_2} &=& \frac{1}{\sqrt{N_1-1}}\sum_{v_1\neq s}\ket{r, v_1},\\
    \ket{\phi_3} &=& \frac{1}{\sqrt{N_2-1}}\sum_{v_2\neq r}\ket{v_2, s},\\
    \ket{\phi_4} &=& \frac{1}{\sqrt{(N_1-1)(N_2-1)}}\sum_{\substack{v_1\neq s\\ v_2\neq r}}\ket{v_2,v_1}.
\end{eqnarray}
The walk is bipartite, then it suffices to analyze the action of $U_{s,r}^2$ in this subspace,
\begin{eqnarray}
    U_{s,r}^2\ket{\phi_1} &=& \frac{(N_1-2)(N_2-2)}{N_1N_2}\ket{\phi_1}+\frac{2\sqrt{N_1-1}(N_2-2)}{N_1N_2}\ket{\phi_2}-\frac{2(N_1-2)\sqrt{N_2-1}}{N_1N_2}\ket{\phi_3}-\nonumber\\
    &&-\frac{4\sqrt{N_1-1}\sqrt{N_2-1}}{N_1N_2}\ket{\phi_4},\\
    U_{s,r}^2\ket{\phi_2} &=& -\frac{2\sqrt{N_1-1}(N_2-2)}{N_1N_2}\ket{\phi_1}+\frac{(N_1-2)(N_2-2)}{N_1N_2}\ket{\phi_2}+\nonumber\\
    &&+\frac{4\sqrt{N_1-1}\sqrt{N_2-1}}{N_1N_2}\ket{\phi_3}-\frac{2(N_1-2)\sqrt{N_2-1}}{N_1N_2}\ket{\phi_4},\\
    U_{s,r}^2\ket{\phi_3} &=& \frac{2(N_1-2)\sqrt{N_2-1}}{N_1N_2}\ket{\phi_1}+\frac{4\sqrt{N_1-1}\sqrt{N_2-1}}{N_1N_2}\ket{\phi_2}+\frac{(N_1-2)(N_2-2)}{N_1N_2}\ket{\phi_3}+\nonumber\\
    &&+\frac{2\sqrt{N_1-1}(N_2-2)}{N_1N_2}\ket{\phi_4},\\
    U_{s,r}^2\ket{\phi_4} &=& -\frac{4\sqrt{N_1-1}\sqrt{N_2-1}}{N_1N_2}\ket{\phi_1}+\frac{2(N_1-2)\sqrt{N_2-1}}{N_1N_2}\ket{\phi_2}-\nonumber\\
    &&-\frac{2\sqrt{N_1-1}(N_2-2)}{N_1N_2}\ket{\phi_3}+\frac{(N_1-2)(N_2-2)}{N_1N_2}\ket{\phi_4}.
\end{eqnarray}
Therefore, $\{\ket{\phi_i}:i=1..4\}$ is an invariant subspace of $U_{s,r}^2$. By considering $\cos\theta_j = 1-\frac{2}{N_j}$ and $\sin\theta_j = \frac{2}{N_j}\sqrt{N_j-1}$ for $j\in\{1,2\}$, we can express $U_{s,r}^2$ in the invariant basis as
\begin{equation}
    U_{red}^2 = \matrx{
        \cos\theta_1\cos\theta_2 & -\sin\theta_1\cos\theta_2 & \cos\theta_1\sin\theta_2 & -\sin\theta_1\sin\theta_2\\
        \sin\theta_1\cos\theta_2 & \cos\theta_1\cos\theta_2 & \sin\theta_1\sin\theta_2 & \cos\theta_1\sin\theta_2\\
        -\cos\theta_1\sin\theta_2 & \sin\theta_1\sin\theta_2 & \cos\theta_1\cos\theta_2 & -\sin\theta_1\cos\theta_2\\
        -\sin\theta_1\sin\theta_2 & -\cos\theta_1\sin\theta_2 & \sin\theta_1\cos\theta_2 & \cos\theta_1\cos\theta_2
    }.
\end{equation}
Diagonalizing $U^2_{red}$ we obtain that its eigenvalues are $\lambda = \{e^{\pm i\alpha},e^{\pm i\beta}\}$, where $\alpha = \theta_1+\theta_2$ and $\beta = \theta_1-\theta_2$. Their associated eigenvectors are
\begin{equation}
    \ket{\pm\alpha} = \frac{1}{2}\matrx{
            1\\
            \mp i\\
            \pm i\\
            1
    }, \ket{\pm\beta} = \frac{1}{2}\matrx{
            -1\\
            \pm i\\
            \pm i\\
            1
    }.
\end{equation}

The initial state (\ref{eq:psi0}) of the algorithm cannot be expressed in the invariant basis. We have to apply the evolution operator one time in order to obtain the walker at $V_2$. 
Then, we have
\begin{equation}
    \ket{\psi(1)} = U_{s,r}\ket{\psi(0)} = -\frac{1}{\sqrt{N_2}}\left(\ket{\phi_1}+\sqrt{N_2-1}\ket{\phi_3}\right).
\end{equation}
In the eigenbasis of $U_{red}^2$, we can express
\begin{equation}
    \ket{\psi(1)} = \frac{1}{2\sqrt{N_2}}\left[\left(1+ i\sqrt{N_2-1}\right)\left(\ket{-\beta}-\ket{+\alpha}\right)+\left(1- i\sqrt{N_2-1}\right)\left(\ket{+\beta}-\ket{-\alpha}\right)\right].
\end{equation}
The state after $2t+1$ steps is
\begin{eqnarray}
    \ket{\psi(2t+1)} &=& U_{s,r}^{2t}\ket{\psi(1)} =\nonumber\\
    &=&\frac{1}{2\sqrt{N_2}}\left[\left(1+ i\sqrt{N_2-1}\right)\left(e^{-i\beta t}\ket{-\beta}-e^{i\alpha t}\ket{+\alpha}\right)+\right.\nonumber\\
    &&+\left.\left(1- i\sqrt{N_2-1}\right)\left(e^{i\beta t}\ket{+\beta}-e^{-i\alpha t}\ket{-\alpha}\right)\right].
    \label{eq:psit}
\end{eqnarray}

The target state is given by
\begin{equation}
    \ket{target} = \frac{1}{\sqrt{N_1}}\sum_{v_1}\ket{r,v_1} = \frac{1}{\sqrt{N_1}}\left(\ket{\phi_1}+\sqrt{N_1-1}\ket{\phi_2}\right).
    \label{eq:target}
\end{equation}
Calculating the fidelity between $\ket{\psi(2t+1)}$ and the desired target state, from~(\ref{eq:psit}) and (\ref{eq:target}), we obtain
\begin{eqnarray}
    \mathcal{F}_1(2t+1) &=& |\braket{\psi(2t+1)|target}|^2\nonumber\\
    &=&\frac{1}{N_1N_2}\left[\left(\sqrt{N_1-1}\sin(\theta_1t)-\cos(\theta_1t)\right)\times\right.\nonumber\\
    &&\times\left.\left(\sqrt{N_2-1}\sin(\theta_2t)-\cos(\theta_2t)\right)\right]^2.
    \label{eq:fid-diff}
\end{eqnarray}
For achieving state transfer with the highest possible fidelity we must choose the number of steps to be the closest odd integer to the time when the fidelity reaches its maximum.  

\begin{figure}[!htb]
\centering
\begin{subfigure}[t]{0.48\textwidth}
         \centering
\includegraphics[scale=0.5]{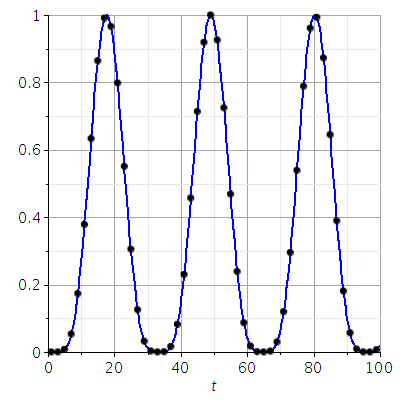}
\caption{$N_1=100, N_2=100$}
\label{fig:fid-diff-100}
\end{subfigure}\hfill
\begin{subfigure}[t]{0.48\textwidth}
         \centering
\includegraphics[scale=0.5]{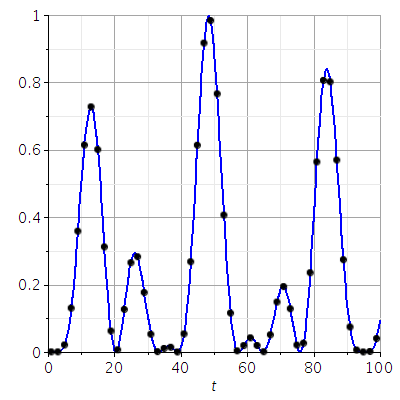}
\caption{$N_1=100, N_2=35$}
\label{fig:fid-diff-100-35}
\end{subfigure}
\caption{The fidelity $\mathcal{F}_1$ in~(\ref{eq:fid-diff}) represented by the blue curve. The black dots are the values for the fidelity at odd steps (at even steps, the fidelity is zero). In (a) $N_1=N_2=100$, the first maximum value $1.0$ occurs at $t\approx17.68$. In (b) $N_1=100, N_2=35$, the maximum in the interval is approximately 0.9952 at $t\approx48.45$.}
\label{fig:fid-diff}
\end{figure}

The fidelity~(\ref{eq:fid-diff}) is depicted in figure~\ref{fig:fid-diff} in the cases $N_1=N_2=100$ (\ref{fig:fid-diff-100}) and $N_1=100,  N_2=35$ (\ref{fig:fid-diff-100-35}). When the partitions have the same size, the fidelity has a nice behavior with peaks reaching 1. The first maximum occurs at $t\approx17.68$, in this case. For the algorithm we set $t=17$ steps, then $\mathcal{F}(17) = 0.9907$.  When the partitions have different sizes, the fidelity has peaks of different sizes. The maximum in the shown interval is $\approx0.9952$ achieved at $t\approx48.45$. For the algorithm we choose $t=49$ steps, then $\mathcal{F}(49) = 0.9835$. 

By making $N_1 = N_2$, the fidelity simplifies to
\begin{equation}
    \mathcal{F}_1(2t+1) = \frac{1}{N_1^2}\left(\sqrt{N_1-1}\sin(\theta_1t)-\cos(\theta_1t)\right)^4.
\end{equation}
For this case,  the first maximum occurs at
\begin{equation}
    T_{max} = 2\left(\frac{\pi-\arctan\sqrt{N_1-1}}{\theta_1}\right)+1,
\end{equation}
and (almost) perfect state transfer is achieved by choosing the number of steps as the odd integer closest to $T_{max}$.

Štefaňák and Skoupý~\cite{Stefanak:2017} analyzed state transfer in the same setting using  the coin operator as $G$ on the non-marked vertices and $-I$ on the marked vertices. We refer to it as the $(G,-I)$ case. We  will compare it with our $(G,-G)$ setting.

\subsection{Comparison between fidelities of $(G,-G)$ and $(G,-I)$ cases}
For the $(G,-I)$~\cite{Stefanak:2017} case with $s$ and $r$ in different partitions, the fidelity is given by
\begin{equation}
    \begin{split}
     \mathcal{F}_2(2t+1) =& \frac{1}{N_1N_2(N_2+N_1-1)^2}\left[N_1N_2-(N_1-1)(N_2-1)\cos(\omega t)+\right.\\
     &\left.+\sqrt{(N_1-1)(N_2-1)(N_1+N_2-1)}\sin(\omega t)\right]^2,
     \end{split}
     \label{eq:fid2}
\end{equation}
where $\omega = \arccos\left(\frac{N_1N_2-2N_1-2N_2+2}{N_1N_2}\right)$. In this case, it is possible to achieve (almost) perfect state transfer only in the case $N_1=N_2$. 

\begin{figure}[!htb]
\centering
\begin{subfigure}[t]{0.48\textwidth}
         \centering
\includegraphics[scale=0.5]{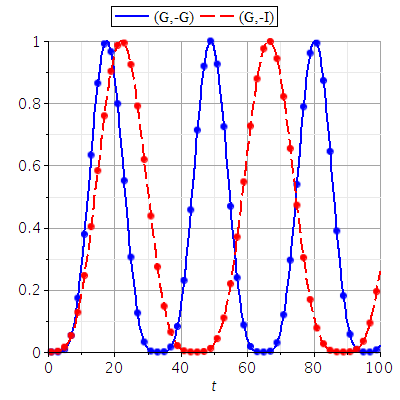}
\caption{$N_1=100, N_2=100$}
\label{fig:fid-comp-100}
\end{subfigure}\hfill
\begin{subfigure}[t]{0.48\textwidth}
         \centering
\includegraphics[scale=0.5]{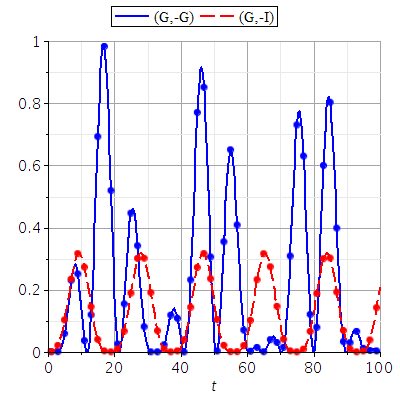}
\caption{$N_1=100, N_2=10$}
\label{fig:fid-comp-100-10}
\end{subfigure}
\caption{The fidelity $\mathcal{F}_1$ depicted by the solid blue curve and $\mathcal{F}_2$ by the dashed red curve. The solid dots represent the values for the fidelity at odd steps (at even steps, the fidelity is zero). In (a) $N_1=N_2=100$, both functions achieve the maximum value of $1.0$ at $t\approx17.68$ and $t\approx22.19$ for $\mathcal{F}_1$ and $\mathcal{F}_2$, respectively.  In (b) $N_1=100, N_2=10$, for $\mathcal{F}_1$ the maximum value of $\approx0.9902$ is obtained at $t\approx 16.74$; for $\mathcal{F}_2$ the maximum value of $\approx0.3180$ is obtained at $t\approx 9.33$.}
\label{fig:fid-comp}
\end{figure}

Now we compare $\mathcal{F}_1$~(\ref{eq:fid-diff}) and $\mathcal{F}_2$~(\ref{eq:fid2}). Figure~\ref{fig:fid-comp} shows the fidelity $\mathcal{F}_1$ in solid blue and $\mathcal{F}_2$ in dashed red. When the partitions have the same size, both functions achieve the maximum value of $1$, as mentioned before. $\mathcal{F}_1$ can achieve the maximum with a fewer number of steps than $\mathcal{F}_2$, which can be observed in~\ref{fig:fid-comp-100} for $N_1=N_2=100$. The biggest difference between the two functions happens when $N_1\gg N_2$ (or the opposite, since the functions are symmetric). State transfer with high fidelity is still possible to be achieved for the case $(G,-G)$ when $N_1 \neq N_2$. For example, in figure~\ref{fig:fid-comp-100-10},  $N_1=100$ and $N_2=10$, the maximum value for $\mathcal{F}_1$ is $\approx0.9902$ at $t\approx 16.74$ and for $\mathcal{F}_2$ the maximum value of $\approx0.3180$ is obtained at $t\approx 9.33$.

\begin{figure}[!htb]
\centering
\begin{subfigure}[t]{0.48\textwidth}
         \centering
\includegraphics[scale=0.5]{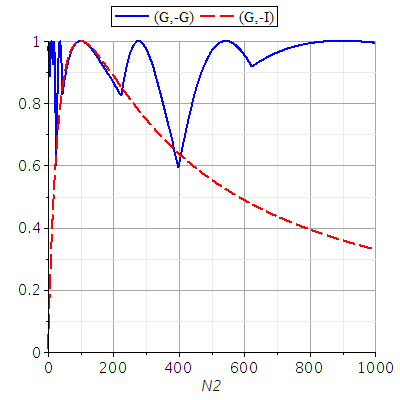}
\caption{Maximum of fidelity}
\label{fig:fmax-comp}
\end{subfigure}\hfill
\begin{subfigure}[t]{0.48\textwidth}
         \centering
\includegraphics[scale=0.5]{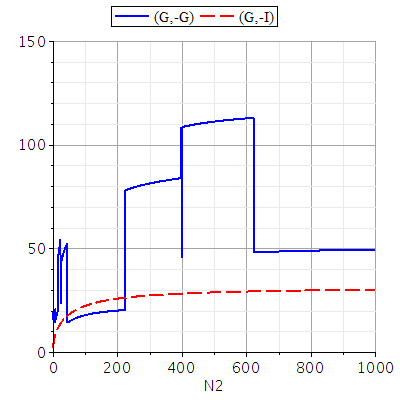}
\caption{Time of maximum}
\label{fig:maxt-comp}
\end{subfigure}
\caption{The (a) maximum value of fidelity $\mathcal{F}_1$ and $\mathcal{F}_2$ and (b) time to achieve it, for  fixed $N_1=100$ and $N_2=\{1,...,1000\}$. The case $(G,-G)$ is depicted by the solid blue curve. The maximum is obtained by using numeric global optimization method in the time interval $(0,5\sqrt{N_1+N_2})$. The case $(G,-I)$ is depicted by the dashed red curve, the expressions for the maximum fidelity and time can be obtained analytically~\cite{Stefanak:2017}.}
\label{fig:max-comp}
\end{figure}

Next, in figure~\ref{fig:max-comp}, we depict the maximum value of the fidelity and the time to achieve it by fixing $N_1=100$ and varying the value of $N_2$ from 1 to 1000. The red dashed curve is the case $(G,-I)$. The expressions for the maximum of $\mathcal{F}_2$ and the time of maximum can be obtained analytically~\cite{Stefanak:2017}. We can observe in~\ref{fig:fmax-comp} how the curve starts to grow from $N_2=1$, reaches 1 at $N_1=N_2$, and starts again to decline for $N_2 > N_1$. For the case $(G,-G)$ depicted by the solid blue curve, we were not able to calculate an analytical expression for the maximum. However, by knowing the values for $N_1$ and $N_2$, we can easily calculate it numerically by using a global optimization method. For this case we obtain the maximum in the time interval $(0,5\sqrt{N_1+N_2})$. We observe we can achieve high fidelity even when $N_1\neq N_2$, but for not all values of $N_2$ in the calculated interval. The number of steps required can be higher for this case.

%%
%% Sender and receiver in the same partition
%%
\section{Sender and receiver in the same partition}
\label{sec:same}
\begin{figure}[!htb]
\centering
\begin{tikzpicture}[thick,
  every node/.style={draw,circle},
  fsnode/.style={fill=black!15},
  ssnode/.style={fill=black!15},
  every fit/.style={ellipse,draw,inner sep=-2pt,text width=2cm},
  -,shorten >= 0pt,shorten <= 0pt
]

%Braces
\begin{scope}[every node/.style={}]
\draw [decorate, 
    decoration = {calligraphic brace,
        raise=5pt,
        amplitude=5pt}] (-0.35,0.1) --  (0.35,0.1)
    node[pos=0.5,above=10pt,black]{$V_1$};
\draw [decorate, 
    decoration = {calligraphic brace,
        raise=5pt,
        amplitude=5pt}] (1.35,-0.4) --  (2.05,-0.4)
    node[pos=0.5,above=10pt,black]{$V_2$};
\end{scope}

% the vertices of U
\begin{scope}[start chain=going below,node distance=4mm]
\node[double,fill=green!15,on chain,minimum size=12pt,inner sep=0pt] (f1) {\footnotesize{$s$}};
\node[double,fill=red!15,on chain,minimum size=12pt,inner sep=0pt] (f2) {\footnotesize{$r$}};
\foreach \i in {3,...,4}
  \node[fsnode,on chain,minimum size=12pt,inner sep=0pt] (f\i) {};

\end{scope}

% the vertices of V
\begin{scope}[xshift=1.7cm,yshift=-0.5cm,start chain=going below,node distance=4mm]
\foreach \i in {1,...,3}
  \node[ssnode,on chain,minimum size=12pt,inner sep=0pt] (s\i) {};
\end{scope}

% the edges
\draw (f1) -- (s1);
\draw (f1) -- (s2);
\draw (f1) -- (s3);
\draw (f2) -- (s1);
\draw (f2) -- (s2);
\draw (f2) -- (s3);
\draw (f3) -- (s1);
\draw (f3) -- (s2);
\draw (f3) -- (s3);
\draw (f4) -- (s1);
\draw (f4) -- (s2);
\draw (f4) -- (s3);
\end{tikzpicture}

\caption{A complete bipartite graph with the sender vertex $s$ and receiver vertex $r$ in the same partition $V_1$. In this case, $N_1 = 4$ and $N_2=3$. }
\label{fig:bip-same}
\end{figure}
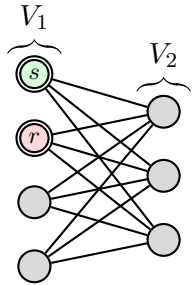

Now consider the sender and the receiver in the same partition, that is, $r, s \in V_1$. See an example in figure~\ref{fig:bip-same}.

As before, we apply $-G$ to the marked vertices and $G$ to the non-marked, that is, $C_s = -G_s$, $C_r = -G_r$, $C_{v_1} = G_{v_1}(v_1\neq\{s,r\})$, $C_{v_2} = G_{v_2}$. The quantum walk is driven by the evolution operator $U_{s,r} = SC$.

Again we can find an invariant subspace in which the quantum walk takes place. Since the receiver vertex $r$ belongs to the first partition, we consider the states that contains vertices of $V_1$ pointing to $V_2$, that is,
\begin{eqnarray}
    \ket{\phi_1} &=& \frac{1}{\sqrt{N_2}}\sum_{v_2}\ket{s, v_2},\\
    \ket{\phi_2} &=& \frac{1}{\sqrt{N_2}}\sum_{v_2}\ket{r, v_2},\\
    \ket{\phi_3} &=& \frac{1}{\sqrt{N_2(N_1-2)}}\sum_{\substack{v_1\neq \{s,r\}\\ v_2}}\ket{v_1,v_2}.
\end{eqnarray}
The initial state for this case is exactly $\ket{\psi(0)} = \ket{\phi_1}$ and the target state $\ket{target} = \ket{\phi_2}$.
In this subspace, $U_{s,r}^2$ acts as 
\begin{eqnarray}
    U_{s,r}^2\ket{\phi_1} &=& \left(1-\frac{2}{N_1}\right)\ket{\phi_1}-\frac{2}{N_1}\ket{\phi_2}-\frac{2\sqrt{N_1-2}}{N1}\ket{\phi_3},\\
    U_{s,r}^2\ket{\phi_2} &=& -\frac{2}{N_1}\ket{\phi_1}+\left(1-\frac{2}{N_1}\right)\ket{\phi_2}-\frac{2\sqrt{N_1-2}}{N1}\ket{\phi_3},\\
    U_{s,r}^2\ket{\phi_3} &=& \frac{2\sqrt{N_1-2}}{N1}\ket{\phi_1}+\frac{2\sqrt{N_1-2}}{N1}\ket{\phi_2}+\left(1-\frac{4}{N_1}\right)\ket{\phi_3}.
\end{eqnarray}
Notice that its behavior is independent of the size of $V_2$. This gives us the same result as in case of the algorithm that uses $(G,-I)$~\cite{Stefanak:2017}. The reduced evolution operator $U^2_{r}$ is also the same as for the algorithm on the star graph~\cite{Stefanak:2016}. From~\cite{Stefanak:2016}, the fidelity is given by
\begin{equation}
\mathcal{F}(2t) = |\braket{\psi(2t)|target}|^2 =  \sin^4\left(\frac{\omega t}{2}\right),
\label{eq:fid-same}
\end{equation}
where 
\begin{equation}
    \omega = \arccos\left(1-\frac{4}{N_1}\right).
\end{equation}

\begin{figure}[!htb]
    \centering
    \includegraphics[scale=0.5]{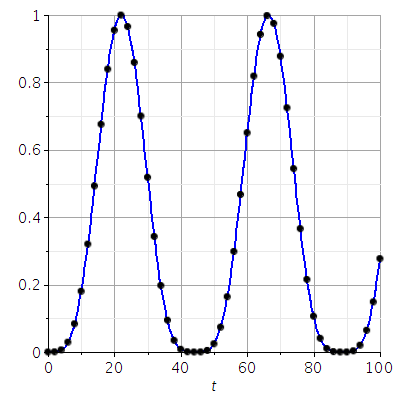}
    \caption{The fidelity (\ref{eq:fid-same}) for $N_1=100$. The black dots are the values for the fidelity at even steps (at odd steps, the fidelity is zero). The maximum value of $1$ is achieved at $t\approx 22.14$. Then, for the algorithm we choose $t=22$ steps, which gives us $\mathcal{F}(22) \approx 0.9998$.}
    \label{fig:fid-same}
\end{figure}

We plot the fidelity for $N_1 = 100$ in figure~\ref{fig:fid-same}. The maximum value of $1$ for the fidelity is achieved at $2\pi/\omega$. Then, the number of steps of the algorithm is the closest even integer to $2\pi/\omega$ in order to achieve state transfer with the highest fidelity possible.

%%
%% Active switch approach
%%
\section{Active switch approach}
\label{sec:switch}

\v{S}tefa\v{n}\'ak and Skoup\'y~\cite{Stefanak:2021} proposed a state transfer algorithm with an active switch strategy based on a single marked vertex. At the beggining, the marked vertex is set as the sender vertex and after some number of steps, the marked vertex is switched to the receiver vertex. Their algorithm is applied to large complete $M$-partite graphs with $N$ vertices in each partition by using lackadaisical quantum walks with self-loop weight equal to 1. Therefore, the complete bipartite graph with $N_1=N_2$ is included in it. Following, we show that a similar active switch strategy also works when $N_1 \neq N_2$.

\subsection{LQW on the complete bipartite graph}

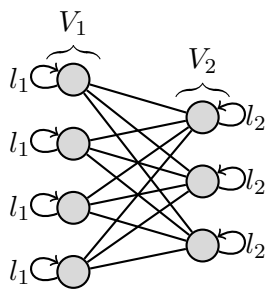
\begin{figure}[!htb]
\centering

\centering
\begin{tikzpicture}[thick,
  every node/.style={draw,circle},
  fsnode/.style={fill=black!15},
  ssnode/.style={fill=black!15},
  every fit/.style={ellipse,draw,inner sep=-2pt,text width=2cm},
  -,shorten >= 0pt,shorten <= 0pt
]

%Braces
\begin{scope}[every node/.style={}]
\draw [decorate, 
    decoration = {calligraphic brace,
        raise=5pt,
        amplitude=5pt}] (-0.35,0.1) --  (0.35,0.1)
    node[pos=0.5,above=10pt,black]{$V_1$};
\draw [decorate, 
    decoration = {calligraphic brace,
        raise=5pt,
        amplitude=5pt}] (1.35,-0.4) --  (2.05,-0.4)
    node[pos=0.5,above=10pt,black]{$V_2$};
\end{scope}

% the vertices of U
\begin{scope}[start chain=going below,node distance=4mm]
%\node[double,fill=green!15,on chain,minimum size=12pt,inner sep=0pt] (f1) {\footnotesize{$s$}};
%\node[double,fill=red!15,on chain,minimum size=12pt,inner sep=0pt] (f2) {\footnotesize{$r$}};
\foreach \i in {1,...,4}
  \node[fsnode,on chain,minimum size=12pt,inner sep=0pt] (f\i) {};

\end{scope}

% the vertices of V
\begin{scope}[xshift=1.7cm,yshift=-0.5cm,start chain=going below,node distance=4mm]
\foreach \i in {1,...,3}
  \node[ssnode,on chain,minimum size=12pt,inner sep=0pt] (s\i) {};
\end{scope}

% the edges
\draw (f1) -- (s1);
\draw (f1) -- (s2);
\draw (f1) -- (s3);
\draw (f2) -- (s1);
\draw (f2) -- (s2);
\draw (f2) -- (s3);
\draw (f3) -- (s1);
\draw (f3) -- (s2);
\draw (f3) -- (s3);
\draw (f4) -- (s1);
\draw (f4) -- (s2);
\draw (f4) -- (s3);

\tikzset{every loop/.style={min distance=5mm,in=150,out=200,looseness=7}, every node/.style={}}

\draw[->]  (f1) edge [loop below] (f1) node[left=12pt,black] {$l_1$};

\draw[->]  (f2) edge [loop below] (f2) node[left=12pt,black] {$l_1$};
\draw[->]  (f3) edge [loop below] (f3) node[left=12pt,black] {$l_1$};
\draw[->]  (f4) edge [loop below] (f4) node[left=12pt,black] {$l_1$};

\tikzset{every loop/.style={min distance=5mm,in=30,out=-20,looseness=7}}

\draw[->]  (s1) edge [loop below] (s1) node[right=12pt,black] {$l_2$};
\draw[->]  (s2) edge [loop below] (s2) node[right=12pt,black] {$l_2$};
\draw[->]  (s3) edge [loop below] (s3) node[right=12pt,black] {$l_2$};

\end{tikzpicture}
\caption{A complete bipartite graph with self-loops. The self-loop weights are $l_1$ and $l_2$ for the vertices in set $V_1$ and $V_2$, respectively.}
\label{fig:lqw}
\end{figure}

Lackadaisical quantum walks are defined by having a weighted self-loop attached to each vertex~\cite{Wong:2017}. In our case, we set different weights for different vertices. The self-loop weight of the vertices in $V_1$ is $l_1$ and the self-loop weight of the vertices in $V_2$ is $l_2$. See figure~\ref{fig:lqw}. 
The quantum walk is a discrete-time coined quantum walk (similar as in section~\ref{sec:QW}), where the evolution operator is given by $U = SC$. The shift operator $S$ is the flip-flop shift~(\ref{eq:shift}). The coin operator is the generalized Grover diffusion~\cite{Wong:2017} given by
\begin{equation}
    \hat{G}_v = \ket{v}\bra{v}\otimes(2\ket{\hat{\mu}_v}\bra{\hat{\mu}_v}-I),
\end{equation}
where
\begin{equation}
    \ket{\hat{\mu}_v} = \frac{1}{d_v}\left(\sum_{u\sim v}\ket{u}+\sqrt{l_v}\ket{v}\right).
\end{equation}
Notice that the degree of each vertex is the number of unweighted edges plus the value of its self-loop weight, that is, $d_{v_1} = N_1+l_1$ and $d_{v_2} = N_2+l_2$.

Let us consider the case of one marked vertex $m$. The coin operator is $C_v = \hat{G}_v (v\neq m)$ and $C_m = -\hat{G}_m$.  Without loss of generality, let the marked vertex $m \in V_1$. As we did in the previous sections, we can find an invariant subspace
\begin{eqnarray}
    \ket{\phi_1} &=& \ket{m,m},\label{eq:phi1}\\
    \ket{\phi_2} &=& \frac{1}{\sqrt{N_2}}\sum_{v_2}\ket{m, v_2},\\
    \ket{\phi_3} &=& \frac{1}{\sqrt{N_2}}\sum_{v_2}\ket{v_2,m},\\
    \ket{\phi_4} &=& \frac{1}{\sqrt{N_2}}\sum_{v_2}\ket{v_2,v_2},\\
    \ket{\phi_5} &=& \frac{1}{\sqrt{N_2(N_1-1)}}\sum_{v_2,v_1\neq m}\ket{v_2,v_1},\\
    \ket{\phi_6} &=&  \frac{1}{\sqrt{N_2(N_1-1)}}\sum_{v_1\neq m, v2}\ket{v_1,v_2},\\
    \ket{\phi_7} &=& \frac{1}{\sqrt{(N_1-1)}}\sum_{v_1\neq m}\ket{v_1,v_1},\label{eq:phi7}
\end{eqnarray}
and obtain a reduced matrix by expressing $U_m$ in this basis. Although
it is possible to diagonalize the reduced matrix analytically in this case, it  results in quite lengthy expressions. We will use the results obtained by
Rhodes and Wong~\cite{Rhodes:2019}, who analyzed the problem of search for marked vertices by using lackadaisical quantum walks on the complete bipartite graph. They proved that for large $N_1$ and $N_2$, the success probability of the search algorithm for the case of one marked vertex is improved when $l_1 = N_2/2N_1$ (independent of $l_2$) and the
initial state is stationary under the quantum walk.

The stationary state is given by
\begin{eqnarray}
\ket{\sigma} &=& \frac{1}{\sqrt{2N_1N_2+l_1N_1+l_2N_2}}\left[\sum_{v_1}\left(\sqrt{l_1}\ket{v_1,v_1}+\sum_{v_2}\ket{v_1,v_2}\right)+\right.\\
&&\left.+\sum_{v_2}\left(\sqrt{l_2}\ket{v_2,v_2}+\sum_{v_1}\ket{v_2,v_1}\right)  \right]  
\end{eqnarray}
which satisfies $U\ket{\sigma} = \ket{\sigma}$. Notice that $U$ is the evolution operator without marked vertices.

In~\cite{Rhodes:2019}, they obtained the success probability of finding a marked vertex. In our algorithm, we will be interested in the fidelity with the target vertex, which in this case is the loop on the marked vertex, $\ket{target} = \ket{m,m}$. Therefore, we need to calculate 
\begin{equation}
    \mathcal{F}(t) = |\braket{target|U_m^t|\sigma}|^2 = |\braket{m,m|U_m^t|\sigma}|^2.
\end{equation}
For large $N_1$ and $N_2$, we can express $\ket{\sigma}$ as
\begin{equation}
    \ket{\sigma} = \frac{1}{\sqrt{2}}\left(\ket{\phi_4} + \ket{\phi_5}\right).
\end{equation}
In the asymptotic eigenbasis of $U_m$ (which was obtained by~\cite{Rhodes:2019} and can be found in Appendix~\ref{appendix} for this case), we have
\begin{equation}
    \ket{\sigma} = -\frac{1}{\sqrt{2}}\ket{\psi_1}+\frac{1}{2}\ket{\psi_2}+\frac{1}{2}\ket{\psi_3}.
\end{equation}
Then, we obtain
\begin{equation}
    U_m^t\ket{\sigma} = -\frac{1}{\sqrt{2}}\ket{\psi_1}+\frac{1}{2}e^{-i\theta t}\ket{\psi_2}+e^{i\theta t}\frac{1}{2}\ket{\psi_3},
    \label{eq:umt}
\end{equation}
where
\begin{equation}
    \theta = \arcsin\sqrt{\frac{2}{N_1}}.
\end{equation}
From (\ref{eq:umt}) we can easily calculate
\begin{equation}
    \mathcal{F}(t) = |\braket{target|U_m^t|\sigma}|^2 = |\braket{m,m|U_m^t|\sigma}|^2 = \frac{1}{4}(\cos\theta t -1)^2
    \label{eq:fid-lqw}
\end{equation}
Then the fidelity reaches 1 at $t=\pi/\theta$. This is exactly the same amount of time obtained by~\cite{Rhodes:2019} so that the success probability reaches one. This means that in the search algorithm, the probability on the marked vertex is contained in the self-loop.

\subsection{The state transfer algorithm with active switch}

We have seen that we can start at the stationary state and reach the target state after a certain amount of time. Or we can think in the opposite way, since the walk is periodic. That is, we can start in the target state and reach the stationary state after some time. Our quantum state transfer algorithm with active switch is described as follows.

The self-loop weights are set to $l_1 = N_2/2N_1$ and $l_2=N_1/2N_2$.
The initial state of the algorithm is at the sender loop, that is,
\begin{equation}
\ket{\psi(0)} = \ket{s,s}.    
\end{equation}
Then, we apply the quantum walk $U_s$ (the sender vertex is the marked vertex) for $T_1 =
round\left(\frac{\pi}{\theta_s}\right)$ steps, where
\begin{equation}
    \theta_{v_j} = \arcsin\sqrt{\frac{2}{N_j}}.
\end{equation}
At this point the current state 
\begin{equation}
    \ket{\psi(T_1)} = U_s^{T_1}\ket{\psi(0)}
\end{equation}
has high fidelity with the stationary state $\ket{\sigma}$ (from~(\ref{eq:fid-lqw})). Now we switch the marked vertex to the receiver vertex, obtaining the operator $U_r$. The quantum walk is evolved for another $T_2 = round\left(\frac{\pi}{\theta_r}\right)$ steps. At the end, we have the state
\begin{equation}
    \ket{\psi(T_1+T_2)} = U_r^{T_2}\ket{\psi(T_1)} = U_r^{T_2}U_s^{T_1}\ket{\psi(0)},
\end{equation}
which has high fidelity with the target state $\ket{r,r}$.

In this case, differently from what happens in the active switch algorithm of~\cite{Stefanak:2021}, we have to know in which partition the sender and receiver vertex belong to in order to correctly set the number of steps $T_1$ and $T_2$. Also notice that since we have to round the time so the number of steps is an integer,  the  algorithm  succeeds  with  fidelity
approaching unity for large $N_1$ and $N_2$. 

\begin{figure}[!htb]
    \centering
    \begin{subfigure}[t]{0.48\textwidth}
         \centering
\includegraphics[scale=0.5]{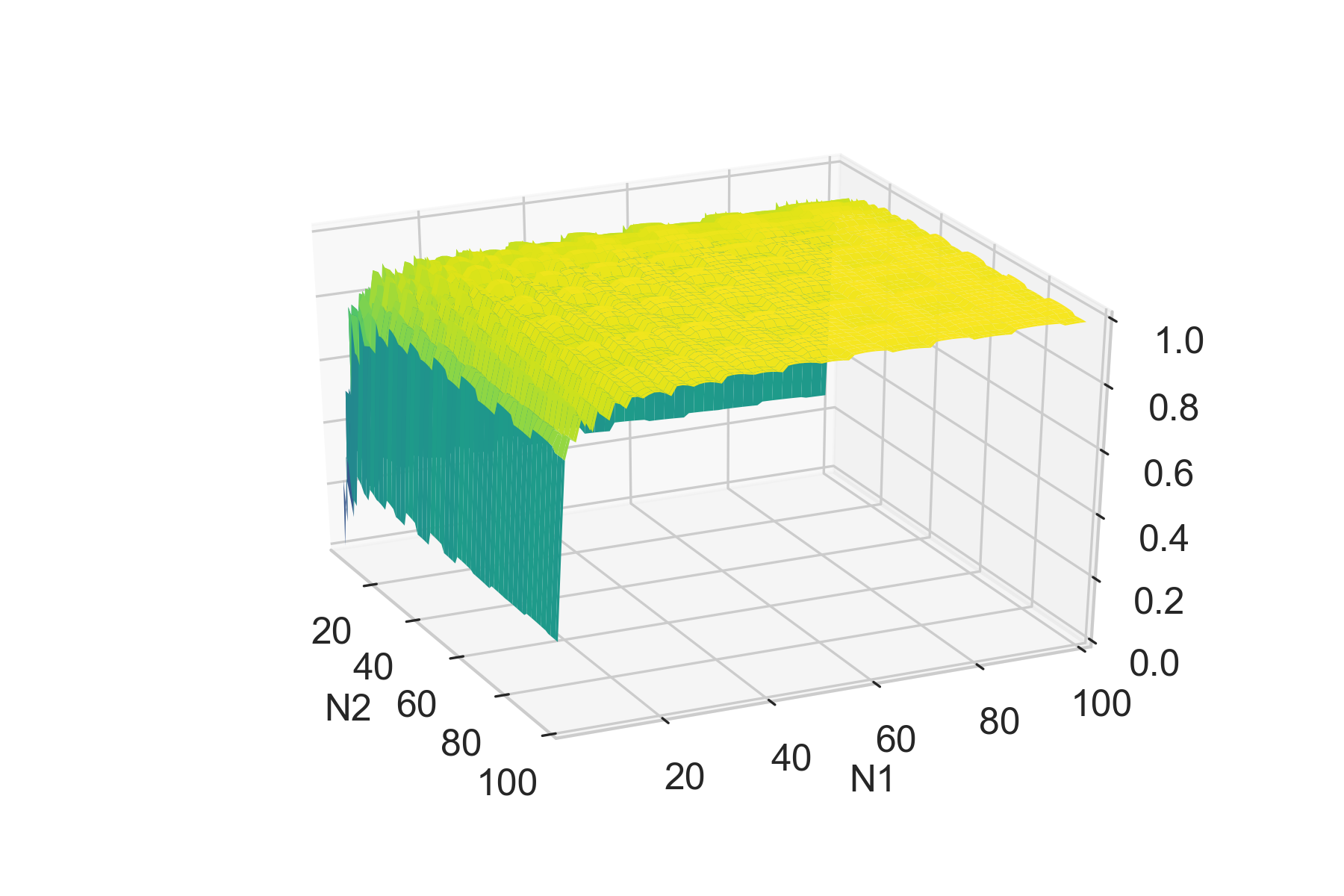}
\caption{$s\in V_1$, $r\in V_2$}
\label{fig:fid3d-diff}
\end{subfigure}\hfill
\begin{subfigure}[t]{0.48\textwidth}
         \centering
\includegraphics[scale=0.5]{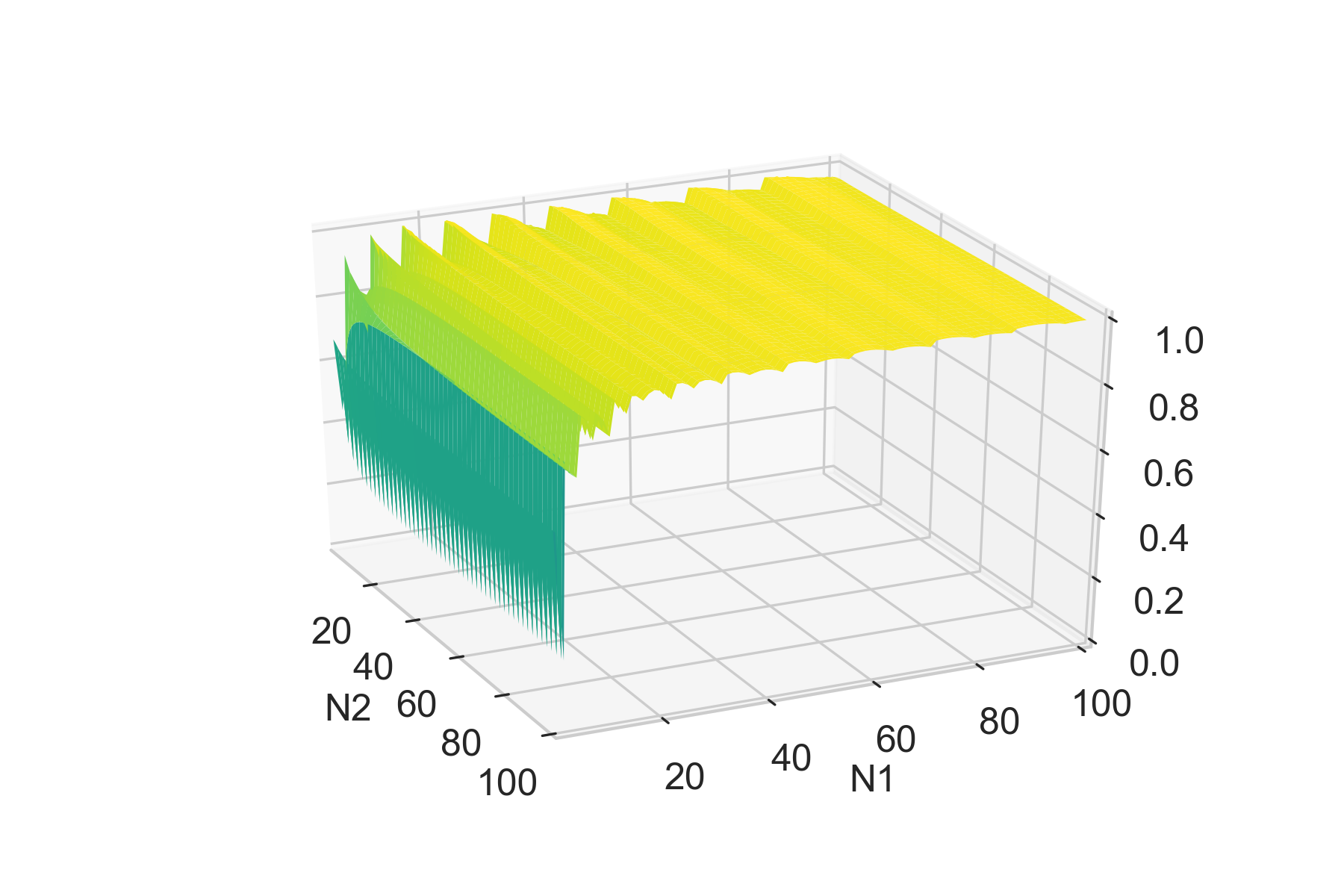}
\caption{$s\in V_1$, $r\in V_1$}
\label{fig:fid3d-same}
\end{subfigure}
    \caption{The fidelity with the target state $\ket{r,r}$ at the end of the state transfer algorithm with active switch. }
    \label{fig:fid-lqw3d}
\end{figure}

\begin{figure}[!htb]
    \centering
    \includegraphics[scale=0.5]{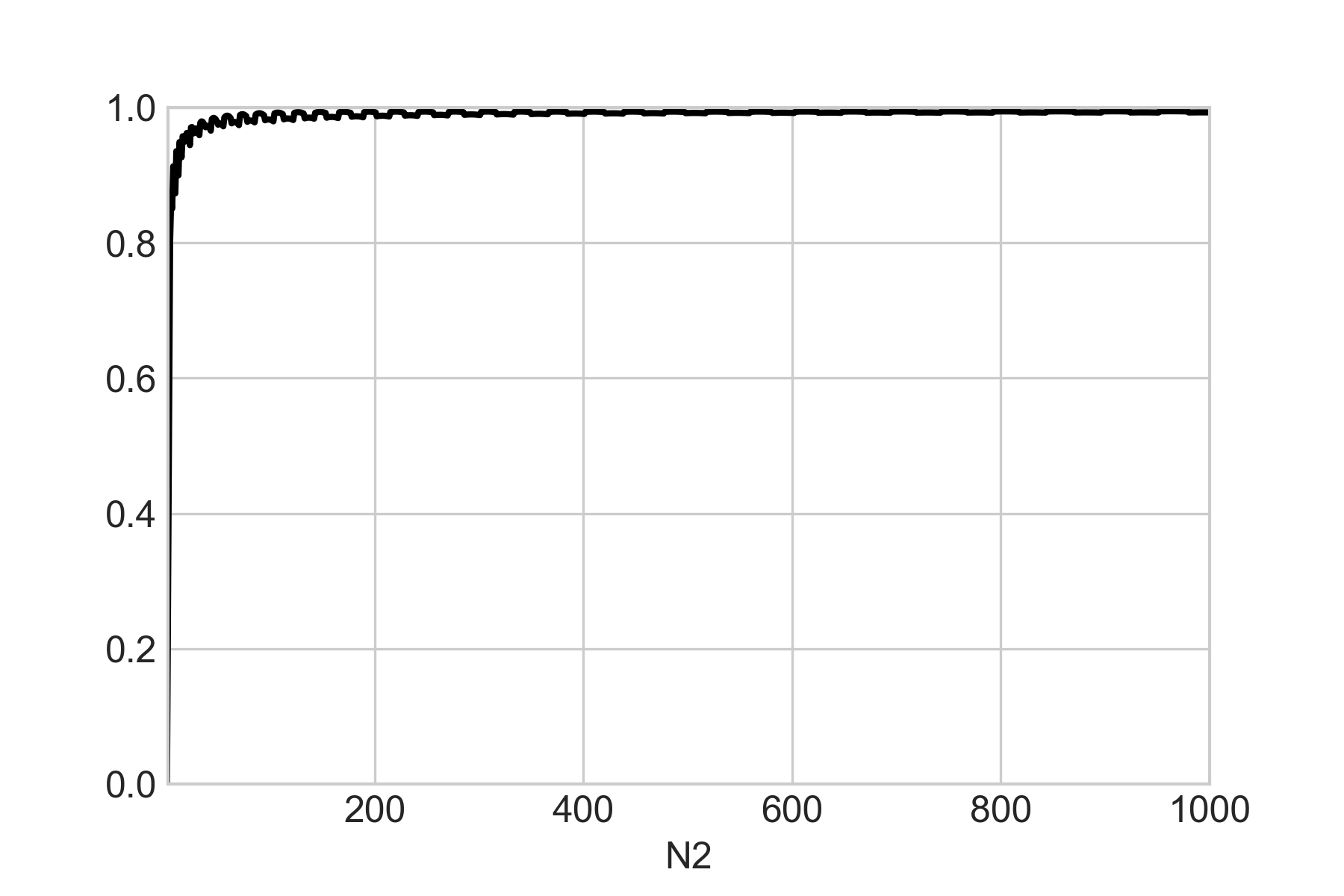}
    \caption{The fidelity with the target state $\ket{r,r}$ at the end of the state transfer algorithm with active switch. In this case, $s\in V_1$ and $r\in V_2$. We set $N_1=100$ and vary $N_2$ in the range $(2,1000)$.}
    \label{fig:fid1000}
\end{figure}

High fidelity state transfer can also be achieved for small graphs. In figure~\ref{fig:fid-lqw3d}, we simulate the algorithm for a certain range of $N_1$ and $N_2$ and we plot the fidelity with the target state $\ket{r,r}$ at the end of the algorithm. In~\ref{fig:fid3d-diff}, we have the sender and receiver in different partitions. In~\ref{fig:fid3d-same}, the sender and receiver are in the same partition. We can observe it is possible to achieve state transfer with high fidelity even for small values of $N_1$ and $N_2$. For example, the value of the fidelity is bigger than 0.9 for $N_1, N_2 > 15$ in both cases, and it will become closer to 1 as $N_1$ and $N_2$ increases. We can observe this fact in figure~\ref{fig:fid1000}, where we fix $N_1 = 100$ and vary $N_2$ in the range $(2,1000)$. In this case, the sender and receiver vertices are in different partitions. 

%%
%% Conclusions
%%
\section{Summary and Conclusions}
\label{sec:conc}
We have analyzed state transfer by using a discrete-time coined quantum walk on the complete bipartite graph. We chose the coin operator as $(G,-G)$, that is, we apply the Grover diffusion coin $G$ to the non-marked vertices and $-G$ to the marked. We were able to analytically calculate the expression for the fidelity with the target state by expressing the evolution operator in a smaller invariant subspace. We have obtained that it is possible to achieve state transfer with high fidelity even in the case where the sender and receiver vertex belong to opposite partitions of different sizes. 

We have compared our result with the case $(G,-I)$~\cite{Stefanak:2017}. When the sender and receiver belong to the same partition, both cases have exactly the same value for the fidelity and they achieve (almost) perfect state transfer. When the sender and receiver are in different partitions, if $N_1 = N_2$, both also achieve (almost) perfect state transfer, with our case requiring less amount of steps. When $N_1 \neq N_2$, our case can achieve high fidelity value for some instances of $N_1$ and $N_2$, specially when $N_1 \gg N_2$ (or $N_2 \gg N_1$), as it was shown numerically. For $(G,-I)$, the value of the fidelity can be quite low in such cases. 

We have presented an alternative solution, which is an active switch approach (similar to the one in~\cite{Stefanak:2021}) that uses lackadaisical quantum walks. By using previous results from~\cite{Rhodes:2019}, we have shown that for large $N_1$ and $N_2$ with proper self-loop weights and considering one marked vertex, if the initial state is the stationary state of the quantum walk, after a certain amount of time the state of the quantum walk will be contained at the self-loop on the marked vertex. With this fact, we have described a quantum state transfer algorithm based on one marked vertex. We start the algorithm by selecting the sender as the marked vertex. The initial state is the sender vertex self-loop. After we evolve the walk for some time, the marked vertex is switched to the receiver vertex. At the end of the algorithm we obtain the walker at the receiver vertex self-loop. Therefore, it is possible to achieve state transfer with  fidelity
approaching unity for a large graph, independent of the location of the sender and receiver vertices. However, in this case, we need to know the location of the sender and receiver vertices in order to correctly set up the number of steps of the algorithm. In addition, through some simulations of the algorithm, we have shown we can achieve state transfer with high fidelity even for small graphs.   
%

%%
%% Acknowledgements
%%
\section*{Acknowledgements}
This work was supported by ERDF project number 1.1.1.2/VIAA/1/16/002.

%%
%% The Bibliography
%%
\bibliographystyle{plainurl}

\begin{thebibliography}{10}

\bibitem{Barr:2014}
Katharine~E. Barr, Tim~J. Proctor, Daniel Allen, and Viv~M. Kendon.
\newblock Periodicity and perfect state transfer in quantum walks on variants
  of cycles.
\newblock {\em Quantum Info. Comput.}, 14(5 \& 6):417–438, 2014.

\bibitem{Bose:2003}
Sougato Bose.
\newblock Quantum communication through an unmodulated spin chain.
\newblock {\em Phys. Rev. Lett.}, 91:207901, 2003.
\newblock
  {\path{doi:10.1103/PhysRevLett.91.207901}}.

\bibitem{Cao:2019}
Wei-Feng Cao, Yu-Guang Yang, Dan Li, Jing-Ru Dong, Yi-Hua Zhou, and Wei-Min
  Shi.
\newblock Quantum state transfer on unsymmetrical graphs via discrete-time
  quantum walk.
\newblock {\em Modern Physics Letters A}, 34(38):1950317, 2019.
\newblock
  {\path{doi:10.1142/s0217732319503176}}.

\bibitem{Chakraborty:2016}
Shantanav Chakraborty, Leonardo Novo, Andris Ambainis, and Yasser Omar.
\newblock Spatial search by quantum walk is optimal for almost all graphs.
\newblock {\em Phys. Rev. Lett.}, 116:100501, 2016.
\newblock 
  {\path{doi:10.1103/PhysRevLett.116.100501}}.

\bibitem{Chen:2019}
Xiu-Bo Chen, Ya-Lan Wang, Gang Xu, and Yi-Xian Yang.
\newblock Quantum network communication with a novel discrete-time quantum
  walk.
\newblock {\em IEEE Access}, 7:13634--13642, 2019.
\newblock
  {\path{doi:10.1109/ACCESS.2018.2890719}}.

\bibitem{Hein:2009}
Birgit Hein and Gregor Tanner.
\newblock Wave communication across regular lattices.
\newblock {\em Phys. Rev. Lett.}, 103:260501, 2009.
\newblock 
  {\path{doi:10.1103/PhysRevLett.103.260501}}.

\bibitem{Stefanak:2016}
M.~\ifmmode \check{S}\else \v{S}\fi{}tefa\ifmmode~\check{n}\else \v{n}\fi{}\'ak
  and S.~Skoup\'y.
\newblock Perfect state transfer by means of discrete-time quantum walk search
  algorithms on highly symmetric graphs.
\newblock {\em Phys. Rev. A}, 94:022301, 2016.
\newblock 
  {\path{doi:10.1103/PhysRevA.94.022301}}.

\bibitem{Kendon:2011}
Vivien~M. Kendon and Christino Tamon.
\newblock Perfect state transfer in quantum walks on graphs.
\newblock {\em Journal of Computational and Theoretical Nanoscience},
  8(3):422--433, 2011.
\newblock 
  {\path{doi:doi:10.1166/jctn.2011.1706}}.

\bibitem{Kurzynski:2011}
Paweł Kurzyński and Antoni Wójcik.
\newblock Discrete-time quantum walk approach to state transfer.
\newblock {\em Physical Review A}, 83(6), 2011.
\newblock 
  {\path{doi:10.1103/physreva.83.062315}}.

\bibitem{Portugal:2013}
R.~Portugal.
\newblock {\em Quantum walks and search algorithms}.
\newblock Springer, New York, 2013.

\bibitem{Rhodes:2019}
Mason~L. Rhodes and Thomas~G. Wong.
\newblock Search by lackadaisical quantum walks with nonhomogeneous weights.
\newblock {\em Phys. Rev. A}, 100:042303, 2019.
\newblock 
  {\path{doi:10.1103/PhysRevA.100.042303}}.

\bibitem{Shang:2019}
Yun Shang, Yu~Wang, Meng Li, and Ruqian Lu.
\newblock Quantum communication protocols by quantum walks with two coins.
\newblock {\em {EPL} (Europhysics Letters)}, 124(6):60009, 2019.
\newblock 
  {\path{doi:10.1209/0295-5075/124/60009}}.

\bibitem{Stefanak:2021}
S.~Skoup\'y and M.~\ifmmode \check{S}\else
  \v{S}\fi{}tefa\ifmmode~\check{n}\else \v{n}\fi{}\'ak.
\newblock Quantum-walk-based state-transfer algorithms on the complete
  $m$-partite graph.
\newblock {\em Phys. Rev. A}, 103:042222, 2021.
\newblock
  {\path{doi:10.1103/PhysRevA.103.042222}}.

\bibitem{Wong:2017}
Thomas~G Wong.
\newblock Coined quantum walks on weighted graphs.
\newblock {\em Journal of Physics A: Mathematical and Theoretical},
  50(47):475301, 2017.
\newblock 
  {\path{doi:10.1088/1751-8121/aa8c17}}.

\bibitem{Yalcinkaya:2015}
İskender Yalçınkaya and Zafer Gedik.
\newblock Qubit state transfer via discrete-time quantum walks.
\newblock {\em Journal of Physics A: Mathematical and Theoretical},
  48(22):225302, 2015.
\newblock 
  {\path{doi:10.1088/1751-8113/48/22/225302}}.

\bibitem{Zhan:2019}
Hanmeng Zhan.
\newblock An infinite family of circulant graphs with perfect state transfer in
  discrete quantum walks.
\newblock {\em Quantum Information Processing}, 18(12):369, 2019.
\newblock 
  {\path{doi:10.1007/s11128-019-2483-3}}.

\bibitem{Zhan:2014}
Xiang Zhan, Hao Qin, Zhi-hao Bian, Jian Li, and Peng Xue.
\newblock Perfect state transfer and efficient quantum routing: A discrete-time
  quantum-walk approach.
\newblock {\em Physical Review A}, 90(1), 2014.
\newblock 
  {\path{doi:10.1103/physreva.90.012331}}.

\bibitem{Stefanak:2017}
M.~Štefaňák and S.~Skoupý.
\newblock Perfect state transfer by means of discrete-time quantum walk on
  complete bipartite graphs.
\newblock {\em Quantum Information Processing}, 16(3):72, 2017.
\newblock 
  {\path{doi:10.1007/s11128-017-1516-z}}.

\end{thebibliography}

\newpage
\appendix

\section{Eigenbasis of $U_m$ for the LQW search on the complete bipartite graph}
\label{appendix}
From~\cite{Rhodes:2019}, considering the invariant subspace given by the states in~(\ref{eq:phi1})-(\ref{eq:phi7}), the evolution operator when searching for one marked vertex $m$ can be written as
\begin{equation}
	U_m = \matrx{
		\frac{N_2-l_1}{N_2+l_1} & \frac{-2\sqrt{N_2l_1}}{N_2+l_1} & 0 & 0 & 0 & 0 & 0 \\
		0 & 0 & \frac{2-N_1-l_2}{N_1+l_2} & \frac{2\sqrt{l_2}}{N_1+l_2} & \frac{2\sqrt{N_1-1}}{N_1+l_2} & 0 & 0 \\
		\frac{-2\sqrt{N_2l_1}}{N_2+l_1} & \frac{l_1-N_2}{N_2+l_1} & 0 & 0 & 0 & 0 & 0 \\
		0 & 0 & \frac{2\sqrt{l_2}}{N_1+l_2} & \frac{l_2-N_1}{N_1+l_2} & \frac{2\sqrt{l_2(N_1-1)}}{N_1+l_2} & 0 & 0 \\
		0 & 0 & 0 & 0 & 0 & \frac{N_2-l_1}{N_2+l_1} & \frac{2\sqrt{N_2l_1}}{N_2+l_1} \\
		0 & 0 & \frac{2\sqrt{N_1-1}}{N_1+l_2} & \frac{2\sqrt{l_2(N_1-1)}}{N_1+l_2} & \frac{N_1-2-l_2}{N_1+l_2} & 0 & 0 \\
		0 & 0 & 0 & 0 & 0 & \frac{2\sqrt{N_2l_1}}{N_2+l_1} & \frac{l_1-N_2}{N_2+l_1}}.
\end{equation}
The spectrum of $U_m$ can be approximated for large $N_1$ and $N_2$. The asymptotic eigenvectors and eigenvalues are
\begin{align}
	\ket{\psi_1}=&\frac{1}{\sqrt{1+\frac{2l_1 N_1}{ N_2}}}\left[1,0,0,0,-\sqrt{\frac{l_1 N_1}{N_2}},-\sqrt{\frac{l_1 N_1}{N_2}},0\right]^\intercal, \> \lambda_1=1, \\
	\ket{\psi_2}=&\frac{1}{\sqrt{2+\frac{N_2}{l_1 N_1}}}\left[1,i\sqrt{\frac{2l_1 N_1+N_2}{4l_1 N_1}},-i\sqrt{\frac{2l_1 N_1+N_2}{4l_1 N_1}},0,\sqrt{\frac{N_2}{4l_1 N_1}},\sqrt{\frac{N_2}{4l_1 N_1}},0\right]^\intercal, \>\lambda_2=e^{-i\theta}, \\
	\ket{\psi_3}=&\frac{1}{\sqrt{2+\frac{N_2}{l_1 N_1}}}\left[1,-i\sqrt{\frac{2l_1 N_1+N_2}{4l_1 N_1}},i\sqrt{\frac{2l_1 N_1+N_2}{4l_1 N_1}},0,\sqrt{\frac{N_2}{4l_1 N_1}},\sqrt{\frac{N_2}{4l_1 N_1}},0\right]^\intercal, \> \lambda_3=e^{i\theta},\\
	\ket{\psi_4}=&\frac{1}{\sqrt{2+\frac{N_2}{l_1 N_1}}}\left[0,1,1,0,0,0,\sqrt{\frac{N_2}{l_1N_1}}\right]^\intercal, \> \lambda_4=-1, \\
	\ket{\psi_5}=&\frac{1}{\sqrt{1+\frac{l_2N_2}{l_1N_1}}}\left[0,0,0,1,0,0,\sqrt{\frac{l_2N_2}{l_1N_1}}\right]^\intercal, \> \lambda_5=-1, \\
	\ket{\psi_6}=&\frac{1}{\sqrt{2+\frac{4(l_1N_1+l_2N_2)}{N_2}}}\left[0,\frac{1}{\sqrt{2}},\frac{1}{\sqrt{2}},\sqrt{2l_2},-i\sqrt{\frac{2l_1N_1+(1+2l_2)N_2}{2N_2}},\right.\nonumber\\
	& \left.i\sqrt{\frac{2l_1N_1+(1+2l_2)N_2}{2N_2}},-\sqrt{\frac{2l_1 N_1}{N_2}}\right]^\intercal, \> \lambda_6=-e^{i\phi},  \\
	\ket{\psi_7}=&\frac{1}{\sqrt{2+\frac{4(l_1N_1+l_2N_2)}{N_2}}}\left[0,\frac{1}{\sqrt{2}},\frac{1}{\sqrt{2}},\sqrt{2l_2},i\sqrt{\frac{2l_1N_1+(1+2l_2)N_2}{2N_2}},\right.\nonumber\\
	&\left. -i\sqrt{\frac{2l_1N_1+(1+2l_2)N_2}{2N_2}},-\sqrt{\frac{2l_1 N_1}{N_2}}\right]^\intercal, \> \lambda_7=-e^{-i\phi},
\end{align}
where 
\begin{gather*}
	\sin \theta=\sqrt{\frac{2l_1 N_1+N_2}{N_1N_2}}, \\
	\sin\phi=\sqrt{\frac{2l_1N_1+N_2+2l_2N_2}{N_1N_2}}.
\end{gather*}

\end{document}